\documentclass[journal]{IEEEtran}

\ifCLASSINFOpdf

\else

\fi
\usepackage[english]{babel}
\usepackage{lipsum}
\usepackage{comment}
\usepackage{nicematrix}
\usepackage{graphicx}
\usepackage{amsmath}
\usepackage{textcomp}
\usepackage{amssymb}
\usepackage[linesnumbered,ruled,vlined]{algorithm2e}
\usepackage{bm}
\usepackage{mathtools}

\usepackage{comment}
\usepackage{textcomp}
\usepackage{microtype}
\usepackage{float}
\usepackage{fancyhdr} % Required for custom headers
\usepackage{xcolor} % Required for text color
\usepackage{booktabs,makecell,tabularx}

\newcolumntype{C}[1]{>{\centering\arraybackslash}p{#1}}
\newcolumntype{L}{>{\raggedright\arraybackslash}X}

\usepackage{siunitx}
\usepackage{etoolbox}
\usepackage{amsthm}
\newtheorem{theorem}{Theorem}[section]

\usepackage{siunitx}
\usepackage{etoolbox}
\usepackage{amsthm}
\newtheorem*{remark}{Remark}
\newrobustcmd{\B}{\bfseries}

\begin{document}

\title{Time-Domain Operational Metrics for Real-time Resilience Assessment in DC Microgrids}

\author{\IEEEauthorblockN{\textsuperscript{}  Maral Shadaei, Ali Hosseinipour, Javad Khazaei}\\
\IEEEauthorblockA{\textit{Department of Electrical and Computer Engineering} \\
\textit{Lehigh University}\\
Bethlehem PA, USA \\
Emails: \textit{masb22@lehigh.edu, alh621@lehigh.edu, khazaei@lehigh.edu}}

\thanks{This research was in part under support from the Department of Defense, Office of Naval Research award number N00014-23-1-2402.}
}

\IEEEpubidadjcol
\IEEEpubid{\begin{minipage}{\textwidth}\ \\[25pt] \centering
  \color{blue}This work has been submitted to the IEEE for possible publication. Copyright may be transferred without notice, after which this version may no longer be accessible.
\end{minipage}}

\maketitle

% As a general rule, do not put math, special symbols or citations
% in the abstract or keywords.
\begin{abstract}
Resilience is emerging as an evolving notion, reflecting a system's ability to endure and adapt to sudden and catastrophic changes and disruptions. This paper spotlights the significance of the quantitative resilience indices of medium-voltage DC (MVDC) distribution technology in marine vessels, notably naval ships. Given the intricate electrical requirements of modern naval ships, the need for a robust power supply underlines the imperative of resilient DC microgrids. Addressing this, our study introduces a novel quantitative metric for operational resilience of DC microgrids based on the measured voltage of main DC bus. This metric not only fuses real-time tracking, compatibility, and computational efficiency, but also adeptly monitors multiple event phases based on time-domain analysis of dc bus voltage dynamics. The intricacies of the dc bus voltage, including overshoots and undershoots, are meticulously accounted for in the algorithm design. With respect to existing research that typically focuses on offline resilience assessments, the proposed index provides valuable real-time information for microgrid operators and identifies whether microgrid resilience is deteriorating over time.
\end{abstract}

% Note that keywords are not normally used for peerreview papers.
\begin{IEEEkeywords}
Quantitative resiliency index,  Medium-voltage DC, Navy microgrids, Real-time resilience tracking.
\end{IEEEkeywords}

% For peer review papers, you can put extra information on the cover
% page as needed:
% \ifCLASSOPTIONpeerreview
% \begin{center} \bfseries EDICS Category: 3-BBND \end{center}
% \fi
%
% For peerreview papers, this IEEEtran command inserts a page break and
% creates the second title. It will be ignored for other modes.
\IEEEpeerreviewmaketitle

\section{Introduction}
% The very first letter is a 2 line initial drop letter followed
% by the rest of the first word in caps.
% 
% form to use if the first word consists of a single letter:
% \IEEEPARstart{A}{demo} file is ....
% 
% form to use if you need the single drop letter followed by
% normal text (unknown if ever used by the IEEE):
% \IEEEPARstart{A}{}demo file is ....
% 
% Some journals put the first two words in caps:
% \IEEEPARstart{T}{his demo} file is ....
% 
% Here we have the typical use of a "T" for an initial drop letter
% and "HIS" in caps to complete the first word.
\IEEEPARstart 
 % While many interpretations and descriptions of resilience have emerged in scholarly works, its exact understanding remains somewhat subtle. 
 Currently, resilience is seen as a evolving idea in power systems, characterized as ‘‘a measure
of the persistence of systems and of their ability to absorb change
and disturbances and recover from failures’’. Broadly, ‘‘resilience is defined
as the ability of equipment, networks, and systems to predict, absorb,
and quickly recover from catastrophic events’’ \cite{RS2}.

Medium-voltage DC (MVDC) distribution technology, used in various infrastructures, offers many benefits, especially in marine vessels such as naval ships. These benefits include easy integration of renewable energy, improved fuel efficiency, and better power quality than AC systems \cite{khazaei2022advances}. Given the significant electrical needs of naval ships, which encompass advanced weaponry, navigation systems, and communication apparatus, MVDC shipboard microgrids are considered in this paper. Due to the significance of mission operation for navy ships, resilience of shipboard microgrids becomes imperative.  

An integral preliminary action to do so, would be to formulate a method to quantitatively assess the operational health and efficacy of the navy's electrical infrastructure on ships \cite{Intro1}. Consequently, it is important that the resilience quantification index possess attributes of compatibility; it should draw insights from historical events and learn from past events, be real-time and online, be applicable at the system level, and be computationally efficient \cite{younesi2022trends}. 

\begin{table*}[tbh] 
\caption{Qualitative comparison between the state-Of-the-art metrics.}\label{tab:comparison}
\centering
\begin{tabular}{@{}llcccc@{}}
\toprule
Ref.                       & Metric                                                                                                                                                                                            & Computationally efficient  & Scalability  & Real-time    & System level \\ \midrule
\cite{55r}                    & $\Re=\xi(L O L P, E D N S, Y, \Psi, \Lambda)$                                                                                                                                                  & $\times$     & $\checkmark$ & $\times$     & Transmission          \\
\cite{9r} & $\begin{aligned} & \mathfrak{R}=\int_t S(t) d t \\ & \mathfrak{R}=\sum n \\ & \mathfrak{R}=\frac{\sum_{i=1}^N t_{u p, i}}{\sum_{i=1}^N\left(t_{u p, i}+t_{\text {down }, i}\right)}\end{aligned}$ & $\times$     & $\times$     & $\times$     & Distribution         \\
\cite{89r}                    & $\vec{\Re}=\left[F_c, D_G, l_G, C_B, C_n, \Lambda_2\right]$                                                                                                                                    &   $\checkmark$           &    $\times$          & $\times$     & Distribution          \\
\cite{77r}                    & $\mathfrak{R}=(\Phi \Lambda E \Pi)$                                                                                                                            &    $\checkmark$          &  $\times$  & $\times$     & Distribution         \\
\cite{58r}                    & $\Re=F I+(1-R E I)+M V I+L L I$                                                                                                                                                           &  $\times$  &  $\times$   & $\checkmark$     & Microgrid          \\
\cite{27r}                    & $\mathfrak{R}=\frac{1}{N T} \sum_{h=1}^{N T} \sum_{t=1}^N \sum_{t=h+1}^{h+H} \Delta t . L S$                                                                                                     &    $\times$  &  $\times$  & $\times$     & Microgrid           \\
\cite{28r}                    & $\Re=\xi(L O L P, E D N S, F, G)$                                                                                                                                                                 &    $\times$     &   $\times$ &  $\checkmark$ & Microgrid            \\
Proposed                   & $\Re= \xi(R_V,V_{DI},V_{REI})$                                                                                                                                                             & $\checkmark$ & $\checkmark$ & $\checkmark$ & Microgrid            \\ \bottomrule
\end{tabular}
\end{table*}

Several studies have focused on developing resilience assessment tools for microgrids and power systems in general \cite{55r,9r,89r,77r,58r,27r,28r}. In Table \ref{tab:comparison}, a comprehensive comparison of various metrics is presented, which compares the application level (i.e., transmission, distribution, and microgrid) and real-time/offline assessment features of existing resilience metrics. As it can be observed, not many existing studies focused on real-time resilience assessment tools for microgrids nor offer a computationally effective approach that is also scalable and can be accommodated for various microgrid designs. To the best of our knowledge, there exists no previous research that presents a set of metrics within the context of real-time operational resilience assessment for DC microgrids that are capable of quantifying various event phases, ensuring computational efficiency, exhibiting compatibility, and offering real-time tracking. To address this gap, we propose a voltage resilience metric in real-time that can be tracked using available measurement and is applicable to various DC microgrid designs. It offers a harmonious blend of compatibility with various microgrid designs, computational efficiency, and real-time applicability, establishing it as a noteworthy breakthrough. Additionally, compared with existing real-time indexes in \cite{58r,28r,Shahideh}, the proposed metric distinguishes itself by providing real-time tracking for various phases of an event in a microgrid, such as the degradation phase, restoration phase, and the degree of degradation.
Specifically, the paper enumerates its distinguished contributions as follows:

 \begin{enumerate}

\item Proposing a distinctive resilience metric tailored for DC microgrids that offers real-time tracking, compatibility, and computational efficiency, filling an evident gap in existing research.
\item The proposed metric is further enhanced to provide real-time monitoring across varied event phases, including the degradation phase, and restoration phase. all contingent on the dc bus voltage dynamics within Navy shipboard microgrids.
\item An algorithm is developed to utilize the proposed resilience index and assess DC microgrid's resilience during events via DC voltage transients and fluctuations.
\end{enumerate}
The rest of the paper is structured as follows. Dynamic model of the system is explained in Section~\ref{sec:MVDC} and the proposed resilience indices are described in Section~\ref{sec:RV}. Section~\ref{sec:results} validates of the proposed metrics through several case studies and conclusions are drawn in Section~\ref{sec:conclusion}.
%-------------------------------------------------------------
\begin{figure}[tbh]
  \centering
    \includegraphics[width=3.3in]{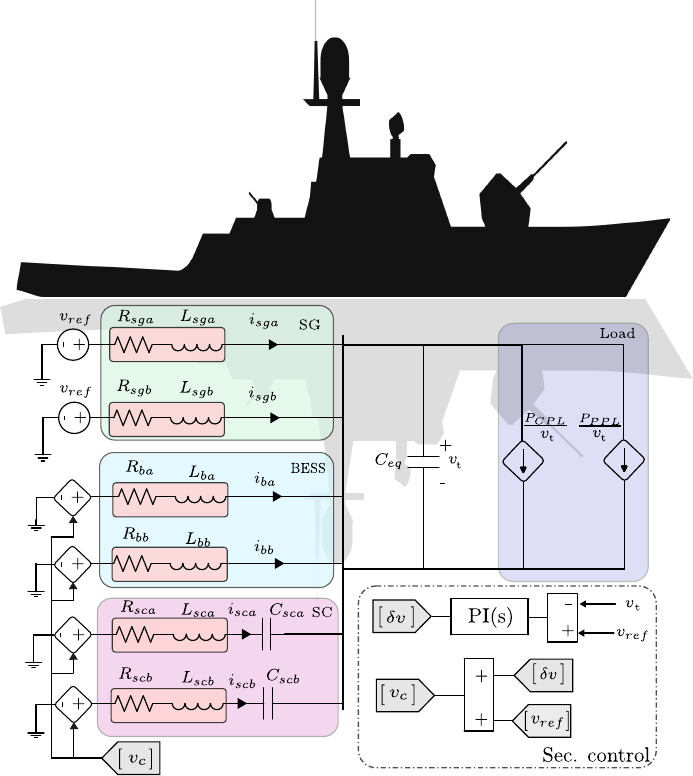}
    \caption{\footnotesize Closed-loop MVDC microgrid configuration for Navy shipboards.}
    \label{MVDC}
  \end{figure}
\section{Methodology} \label{sec:MVDC}
\subsection{Model Description}
Fig.~\ref{MVDC}, illustrates an equivalent circuit of a 6kV MVDC shipboard microgrid inspired from \cite{multiconverter}. Within this microgrid, power sources operate with secondary control design.  Synchronous generators (SGs) mainly contribute to the low-frequency loads,such as constant power loads (\text{$P_{CPL}$}). To address rapidly varying loads such as pulsed power loads (\text{$P_{PPL}$}), two battery energy storage units (BESS) and supercapacitors (SCs) are designed and to reduce the current stress on BESSs. In this system, subsequent to any event or load changes, the droop controller determines voltage shifts for the main bus. This can be mathematically written by:
\begin{align}
 v_t &= {v_{o_{DC}}} - R_{\text{eq}} i_{\text{load}} \\
&  = {v_{o_{DC}}} - \left(R_{s g a}||R_{s g b}|| R_{b a} || R_{b b}\right)\frac{P_{\text{load}}}{v_t}
\end{align}
where ${v_{o_{DC}}}$ stands for the reference voltage which is  6KV in this system. Also, $v_t$ is the MVDC real-time voltage measurement across the DC link capacitor $C_{\text{eq}}$, and $R_{eq}$ denotes the equivalent resistive droop gain of the system for the conventional generators and BESSs.  Comprehensive information regarding the droop control design for the ship can be found in \cite{hosseinipour2023multifunctional}. The total load power is also represented by $P_\text{load}$.
To restore voltage fluctuations arising from the droop controller after events, the secondary voltage control has been utilized to stabilize the voltage trajectory, as illustrated in Fig.~\ref{MVDC}.
\subsection{Reduced order model of MVDC shipboard microgrids}
The open loop differential equations of the system in Fig.~\ref{MVDC} can be formulated as follows \cite{multiconverter}: 
\begin{subequations} \label{dynmvdc-openloop}
\begin{align}
C_{\text{eq}} \dot{v_t} &= \frac{}{}\sum_{i \in \mathcal{N}_{SG}} (i_{\text{SG}i}) +\sum_{i \in \mathcal{N}_{B}} (i_{\text{B}i}) +\sum_{i \in \mathcal{N}_{SC}} (i_{\text{SC}i}) \nonumber \\
&- \frac{P_{\text{CPL}}}{v_t} -\frac{P_{\text{PPL}}}{v_t} \label{eqn:cap_voltage} \\
L_{i} \dot{i}_{i} &= {v_{o_{DC}}} - R_{i}i_{i} - v_t + \delta v_i \quad \forall i \in \{\mathcal{N}_{SG},\mathcal{N}_{B}\} \label{eqn:sg_current} \\
L_{i}\dot{i}_{i} &= {v_{o_{DC}}} - R_{i}i_{i} - v_{SCi} - v_t \quad \forall i \in \mathcal{N}_{SC} \label{eqn:sc_current} \\
C_{i}\dot{v}_{Ci} &= i_{i} \quad \forall i \in \mathcal{N}_{SC} \label{eqn:sc_voltage}
\end{align}
\end{subequations}
where $L_i$ is the equivalent inductance of the $i$-th generation unit, $v_t$ denotes the output voltage in real-time, $i_i$ is the current of the $k$-th generation unit, and $R_{i}$ is the resistive droop gains for the conventional generators and BESSs, respectively. Centralized $\delta v$ using a PI controller is used, which guarantees voltage restoration (Fig.~\ref{MVDC}).

% The the closed-loop state representation of this MG with the PI controller as a secondary control can be expressed as: 
% \begin{subequations} \label{closed-loop eqaution}
% \begin{align} 
% C_{e q} \dot{x}_1 &= \sum_{i=2}^{7} x_i - \left( \frac{u_1 + u_2}{x_1} \right) \\
% L_{i} \dot{x}_{2,3,4,5} &= {v_{o_{DC}}} - R_{\text{i}} x_{2,3,4,5}  - x_1 + k_p ({v_{o_{DC}}} - x_1) \\
% &\quad - x_{10} + x_{11} \quad \forall i \in \{\mathcal{N}_{SG},\mathcal{N}_{B}\}\\
% L_{s c a} \dot{x}_{6} &= {v_{o_{DC}}} - R_{s c a} x_6 - x_8 - x_1 \\
% L_{s c b} \dot{x}_7 &= {v_{o_{DC}}} - R_{s c b} x_7 - x_9 - x_1 \\
% C_{sca} \dot{x}_8 &= x_6 \\
% C_{scb} \dot{x}_9 &= x_7 \\
% \dot{x}_{10} &= k_i {v_{o_{DC}}} \\
% \dot{x}_{11} &= k_i x_1
% \end{align}
% \end{subequations}
% where $\bm{x} \in \mathbb{R}^{n_x}$ = $[v_t \ I_{SGa} \ I_{SGb} \ I_{Ba} \ I_{Bb} \ I_{SCa} \ I_{SCb} \int I_{SCa} \, dt\ \\ \int I_{SCb} \, dt\ \int k_i {v_{o_{DC}}} \, dt \ \int k_i v_t \, dt ]^T$ is the state variable vector, $\bm{u}  \in \mathbb{R}^{n_u} = [P_{CPL} \: \:P_{PPL}]^T$ denotes the control input vector, and $k_i$ and $k_p$ are the coefficients of the PI controller.
\subsection{MG degredation model}
\begin{figure}[tbh]
  \centering  
   \vspace{-0.2cm} \includegraphics[width=3.2in] {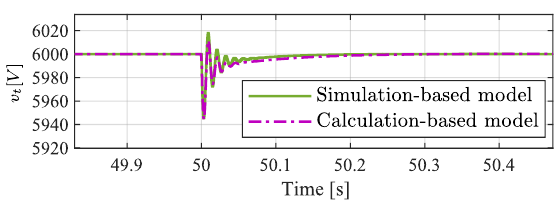} \vspace{-0.05in}
    \caption{\footnotesize Verifying the calculation-based model under sudden load change.} \label{verific}
  \end{figure}

This phase is crucial for assessing the behavior of the MG during both degradation and restoration periods. In Fig. \ref{verific}, we illustrate the voltage behavior of the main bus following an sudden load change. The figure also includes a comparative analysis, contrasting the dynamic behavior of a MATLAB Simulink function consists of reduced order model in the above-mentioned subsection with a detailed simulation model created with MATLAB Simscape library. The findings highlight a notable similarity in dynamic behavior between the reduced-order closed-loop model and the detailed simulation model.

%-------------------------------------------------------------
\section{Proposed Time-domain Voltage Resiliency Index} \label{sec:RV}
When an extreme event impacts the power system, the system's attributes might deviate from their normal state. This time-dependent fluctuation of a system attribute in response to such an event is termed as the system performance. Power system performance can be assessed using various metrics of performance (MoP), including:
\begin{itemize}
    \item [] 1. The percentage of total or critical loads. \item [] 2. The number of supplied critical loads. 
    \item [] 3. The count of survived or failed loads. 
    \item [] 4. Technical metrics, i.e., voltage and frequency \cite{Shahideh}.
\end{itemize} In this study, we underscore the significance of ranking MoPs based on the system's operational strategies and goals in real-time. As such, our primary emphasis is on appraising system's resilience and technical performance via voltage indices. \begin{figure}[b]
\centering
\includegraphics[width=3.2in]{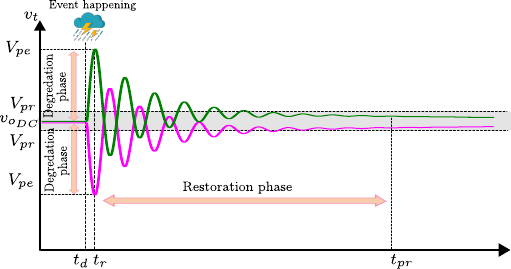}
\caption{\footnotesize Microgrid voltage performance curve during an event.}
\label{Curve}
\end{figure} 

Fig.~\ref{Curve} depicts the voltage response of a DC microgrid during extreme events, such as sudden load changes or generator failures in the MG. Upon the event's onset at $t_d$, the system's performance experiences a noticeable degradation to $V_{pe}$, causing the voltage to deviate from its reference value $v_{o_{DC}}$. This deviation, which may manifest as overshoot or undershoot, reflects the system's transient response to such disturbances. As time progresses to $t_r$, the system begins its restoration phase, facilitated by the MG controllers. By time $t_{pr}$, the system voltage has been fully restored to its reference or desired value. Hence, the proposed set of resiliency metrics, representing various phases of the system during an event is formulated as: 
\begin{align} \label{eq:proposed_metrics}
      \Re= \xi(R_V,V_{DI},V_{REI})
\end{align}

To evaluate the MG's ability to draw insights from past events, and show its degradation over time, we introduce the index \( R_V \). This metric quantifies the MG's voltage compatibility and its capacity to retain memory from prior events. More importantly, it manifests the MG's inherent capability to learn and remember past voltage perturbations.  Therefore, mathematically, it can be formulated as:
\begin{align} \label{eq:RV}
     R_V(t) = \int_{}^{} \left[ {v_{o_{DC}}} - v_t(t) \right] dt 
\end{align}

This equation captures the cumulative discrepancy between the reference voltage ${v_{o_{DC}}}$ and the real-time voltage ${v_t(t)}$ from the disturbance's inception to its termination. To calculate the corrosponding area, we used the trapezoidal rule. 

\begin{theorem}
The trapezoidal rule estimates the integral of a function by approximating the area under its graph as trapezoids. For enhanced accuracy, one can partition the interval into smaller subintervals. Let $\left\{x_k\right\}$ be a partition of $[a, b]$ such that \( a=x_0<\cdots<x_{N-1}<x_N=b \) and \( \Delta x_k=x_k-x_{k-1}  \) be the length of the \( k \)-th subinterval, then the composite trapezoidal rule for uniform subintervals of size \( \Delta x \), is given by:
\begin{align} \label{eq:trap}
\int_a^b f(x) d x \approx \sum_{k=1}^N \frac{f\left(x_{k-1}\right)+f\left(x_k\right)}{2} \Delta x_k
\end{align}
For uniform subintervals of size \( \Delta x \), the formula simplifies to \cite{trap}:
\begin{align} \label{eq:subint}
\int_a^b f(x) d x \approx \frac{\Delta x}{2} \left[f\left(x_0\right)+2 \sum_{k=1}^{N-1} f\left(x_k\right)+f\left(x_N\right)\right]
\end{align}
\end{theorem}

\begin{remark}
    To apply this method in a real-time application, we should update the value of b to correspond only to the next $x_k$ at each simulation step.
\end{remark}

The proposed normalized degradation index (\( V_{DI} \)) is a metric designed to quantitatively assess the resilience of a system in real-time. It is formulated as follows:
\begin{align} \label{eq:DI}
V_{DI}=k*\frac{\int_{t_d}^{t_{r}}({v_{o_{DC}}}-v_t(t)) dt}{{v_{o_{DC}}}(t_{r}-t_d)}
\end{align}
where \( k \) is a scaling coefficient, \( v_{o_{DC}} \) represents the optimal system performance, \( v_t(t) \) is the actual system performance at time \( t \), and \( t_d \) and \( t_r \) are the times of degradation onset and recovery, respectively. The integral in the numerator calculates the total deviation of the system's performance from its optimal state over the specified time period, while the denominator normalizes this deviation. In this index, a value of 0 indicates a highly resilient system with no degradation, while a value of 1 denotes a fragile system that has completely lost functionality. 
\textbf{Algorithm 1} displays the implementation of the $V_{DI}$ index in this study.

 \begin{remark}Upon the occurrence of a disturbance, the voltage signal may exhibit transient behaviors, notably voltage rise or drop. To accurately account for this dynamic response in the proposed ${V_{DI}}$ algorithm, it becomes imperative to detect the value of ${V_{pe}}$  using instantaneous voltage measurements.
 \end{remark}

\begin{algorithm} 
\caption{$V_{DI}$ Calculation Algorithm} \label{algo:DI}
\SetAlgoNlRelativeSize{0}
\KwIn{$\text{clock}, v_t, {v_{o_{DC}}}$}
\textbf{Initialize persistent variables:}
 \hspace{1em} $time, v_m(t), v_m(t-1), t_d, t_r, S_{total},$ 
 \hspace{1em} $Denom, t1, \Delta t $

\If{$time$ is empty}{
    $time \gets 0$
}
:{Initialize other persistent variables similarly}
\BlankLine
\textbf{Execute Nominator calculations}\;
$\Delta v \gets v_t - {v_{o_{DC}}}$\; 

\If{$v_t < {v_{o_{DC}}}$ \textbf{or} $v_t > {v_{o_{DC}}}$}{
    $\delta t \gets \lvert time - clock \rvert$\;
    $time \gets clock$\;
    $S1 \gets \lvert \Delta v \rvert \times \delta t$\; 
    $S_{total} \gets S_{total} + S1$\;

    \textbf{Execute Denominator calculations}\;
    \If{$v_t < {v_{o_{DC}}}$ \textbf{or} $v_t > {v_{o_{DC}}}$}{
        $t1 \gets t1 + 1$\;
        $t_r \gets clock$\;
        \If{$t1 < 1.5$}{
            $t_d \gets clock$\;
        }
        $\Delta t \gets \lvert t_r - t_d \rvert$\;
        $Denom \gets {v_{o_{DC}}} \times \Delta t$\;
    }
}

\If{$v_m(t) == 0$ \textbf{or} ($v_m(t-1) < v_m(t)$ \textbf{and} $v_t < {v_{o_{DC}}}$) \textbf{or} ($v_m(t-1) > v_m(t)$ \textbf{and} $v_t > {v_{o_{DC}}}$)}{
    $t1, t_r, t_d, S_{total} \gets 0$\;
}

\If{$v_m(t) == 0$ \textbf{or} ($v_m(t-1) < v_m(t)$ \textbf{and} $v_t < {v_{o_{DC}}}$) \textbf{or} $v_m(t-1) == v_m(t)$ \textbf{or} ($v_m(t-1) > v_m(t)$ \textbf{and} $v_t > {v_{o_{DC}}}$)}{
    $V_{DI} \gets 0$\;
}
\Else{
    $V_{DI} \gets \frac{S_{total}}{Denom \times {v_{o_{DC}}}}$\;
}

$v_m(t-1) \gets v_m(t)$\;
\Return $V_{DI}$\;

\end{algorithm}
Starting from \( t_r \), the system transitions into its restoration phase, swiftly recovering to an acceptable performance level denoted by \( V_{pr} \) at \( t_{pr} \). This rapid recovery is made possible by the network's redundancy, particularly through the design of an effective secondary controller. To evaluate the efficiency of this restoration phase, we introduce the voltage restoration efficiency index, \( V_{REI} \), defined in Equation~\eqref{eq:REI} as:

\begin{equation} \label{eq:REI}
V_{REI} = \frac{\int_{t_r}^{t_{pr}} (v_t(t) - V_{pe}) dt}{({v_{o_{DC}}} - V_{pe})(t_{pr} - t_r)}
\end{equation}

For an MG that can fully restore its functionality, the metric returns a value of 1. Conversely, if the MG fails to recover, the metric is calculated to be 0. 
Furthermore, this metric signifies how \emph{fast} the MG can bounce back from the particular event.
%%%%%%%%%%%%%%%%%%%%%%%%%%%%%%%%%%%%%%%%5
\section{Simulation results} \label{sec:results}
The proposed resiliency evaluation method is validated for the MVDC microgrid depicted in Fig. 1. The simulation is conducted in MATLAB/Simulink running on a PC with Intel Core i9-10900X 3.7GHz and 64GB RAM under Windows 10. The sampling time for the simulation is set to $50~\mu s$. 
% Next, through simulations involving step changes in load (\text{$P_{CPL}$}) at $t=$\SI{6}{s} and synchronous generation failures at $t=$\SI{10}{s}, we monitor the DC bus voltage to evaluate the resiliency using our proposed indices.
%%%%%%%%%%%%%%%%%%%%%%%%%%%%%%%%%%%%%%%%%%%%%%
\subsection{Case studies}
\begin{figure}[tbh]
  \centering 
   \includegraphics[width=3.2in]{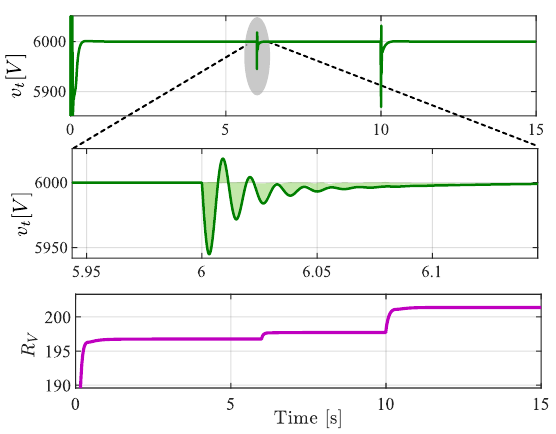} \vspace{-0.1in}
    \caption{\footnotesize DC bus voltage and the $R_V$ index verification.}
    \label{RV_voltage}
  \end{figure}

\begin{table}
\caption{ \footnotesize Different events and area analysis for $R_V$ index}\label{tab:RV}
\resizebox{\columnwidth}{!}{%
\begin{tabular}{@{}ccccc@{}} 
\toprule
 & Time (s) & Area before event & Area after event & Added area \\ \midrule
\begin{tabular}[c]{@{}c@{}}Event 1\\ (\text{$P_{CPL}$} Load Change)\end{tabular}  & t = 6    & 196.7             & 197.7            & 1          \\ \midrule
\begin{tabular}[c]{@{}c@{}}Event 2\\ (Gen. failure)\end{tabular} & t = 10   & 197.7             & 201.1            & 3.4        \\ \bottomrule
\end{tabular}
}
\end{table}

\begin{comment}
The graph of $v_o$ in Fig.~\ref{RV_voltage}  displays how the DC bus voltage reacts to different incidents. Two key events have occured: a sudden change in constant power load (\( P_{CPL} \)) at \( t = 6~s \), and the failure of a synchronous generator at \( t = 10~s \), with the load change emphasized for better visibility. The final part of Fig.~\ref{RV_voltage} validates the ${R_V}$ index, tracking its immediate response to fluctuations in the DC bus voltage. It graphically presents how the ${R_V}$ index evolves, reflecting changes due to the computed area between the actual voltage deviation and the standard voltage according to Equation (\ref{eq:RV}). The ${R_V}$ index visibly increases step by step in response to particular events and stops updating when the voltage stabilizes, offering insights into the system's historical response to disturbances. Table \ref{tab:RV} complements this by detailing the impact of such events on the ${R_V}$ index, categorizing them by time and impact magnitude, which helps in evaluating the system's resilience to disruptions.
\end{comment}
Fig.~\ref{RV_voltage} depicts the DC bus voltage's response to a sudden load change from 10 MW to 15 MW at \( t = 6s \) and a generator failure at \( t = 10s \). The ${R_V}$ index reflexes changes due to the computed area between the actual voltage deviation and the reference voltage according to Equation (\ref{eq:RV}). The ${R_V}$ index visibly increases step by step in response to particular events and stops updating when the voltage stabilizes, offering insights into the system's historical response to disturbances. This graphical evolution of the ${R_V}$ index, alongside Table \ref{tab:RV} that categorizes the impact of such events by time and magnitude, provides a comprehensive analysis of the system's resilience to disruptions.
\begin{comment}

\vspace{-0.1in}
\begin{figure}[tbh]
  \centering
    \includegraphics[width=3.2in]{Figs/VI_inde.pdf} \vspace{-0.1in}
    \caption{\footnotesize Voltage vulnerability index verification.}
    \label{VI}
  \end{figure}

Fig.~\ref{VI} depicts the verification of the voltage vulnerability index (\text{$V_{VI}$}), with $k=0.1$, under specified event scenarios. The \( V_{VI} \) index elucidates how \emph{low} resilience drops when the system encounters perturbations. The consistent monitoring of DC bus voltage and its subsequent deviations provide invaluable insights into the system's vulnerability under these conditions.
    
\end{comment}
  \begin{figure}[tbh]
  \centering
\includegraphics[width=3.2in]{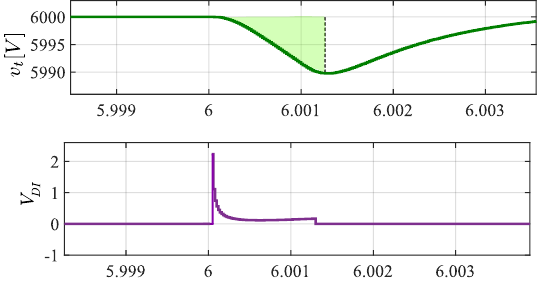} \vspace{-0.1in}
    \caption{ \footnotesize Voltage degradation index verification.}
    \label{DI}
  \end{figure}
  
Fig.~\ref{DI}  presents the verification of the voltage degradation index within a narrowly defined time window. The upper subplot displays the behavior of the DC bus voltage, \( V_t \), capturing a distinct perturbation around \( t = 6s \). The  green area underscores the degradation area of this perturbation over time. Correspondingly, the lower subplot illustrates the ${V_{DI}}$ index's response with $k=10e^{-5}$ to this voltage deviation. The ${V_{DI}}$ index, rendered in purple, signifies the normalized area of the degradation of voltage in real-time. As illustrated, this index not only pinpoints the onset of the event but also discerns the termination of the degradation phase. The precise time scale coupled with the conspicuous shift in the ${V_{DI}}$ index underscores the criticality of immediate monitoring for maintaining voltage stability.
\begin{figure}[tbh]
  \centering
    \includegraphics[width=3.2in]{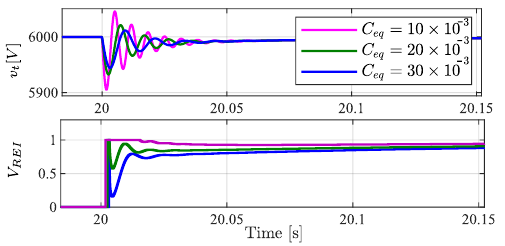}\vspace{-0.1in}
    \caption{\footnotesize Impact of $C_{eq}$ on voltage transient behavior of MVDC and the voltage restoration efficiency index verification.}
    \label{REI}
  \end{figure}

In the MVDC system, the \( C_{eq} \), is indicative of the system's virtual inertia. Larger values of \( C_{eq} \) are associated with a reduced energy imbalance~\cite{dcinertia}. Fig. \ref{REI} provides a clear demonstration of the role that \( C_{eq} \) plays in influencing the transient voltage behavior within the MVDC. This figure also serves to confirm the effectiveness of the \( V_{REI} \) Index. The dynamic response of the \( V_{REI} \) to sudden load changes is captured in the figure, which highlights the effect of different \( C_{eq} \) values on the system's ability to recover post-disturbance. Specifically, the figure quantifies the voltage deviation area as the restoration process begins. It reveals that a smaller \( C_{eq} \) results in a \emph{faster} system recovery time, improving the system's capability to rebound from significant disturbances. An index value approaching 1 denotes a more rapid restoration to normal operational conditions. Moreover, the figure underscores the principle that greater inertia corresponds to increased MG resilience.

\section{Conclusion} \label{sec:conclusion}
This paper introduces innovative quantitative metrics designed to assess the resilience of MVDC microgrids, with a particular emphasis on naval ships. Based on time-domain analysis of DC bus voltage dynamics, three resilience metrics are proposed that include: 1) voltage resilience index \( R_{V} \) to assess the resilience of voltage after events, 2) voltage restoration index \( V_{REI} \)  to assess the resilience of DC microgrid in restoring the voltage to its nominal value after an event, and 3) voltage dip index \( V_{DI} \) to evaluate the strength of an event and its impact on the voltage dip.  The proposed metrics not only showcase real-time tracking capabilities, computational efficiency, and compatibility but also enable operators to monitor the microgrid's recovery speed, the depth of voltage dips, and potential resilience deterioration over time. The validation of these metrics through simulations—encompassing sudden load changes and equipment failures—demonstrates their efficacy and flexibility. Additionally, it has been demonstrated that the voltage recovery index \( V_{REI} \) is tightly related to the inertia of the DC microgrids  \( C_{eq} \) and is capable of accurately tracking the microgrid's recovery speed after an event.
%%%%%%%%%%%%%%%%%%%%%%%%%%%%%%%%%%%%%%%%%%%%%%%%%%%%%%

\bibliography{references.bib}
\bibliographystyle{IEEEtran}

% that's all folks
\end{document}